\documentclass[epj]{webofc}
\usepackage[varg]{txfonts}   
\woctitle{MESON2018 - the 15$^\textrm{th}$ International Workshop on Meson Physics}
\begin{document}
\selectlanguage{english}
\title{A revision of radiative corrections to double-Dalitz decays}

\author{Pablo~Sanchez-Puertas\inst{1}\fnsep\thanks{\email{sanchezp@ipnp.troja.mff.cuni.cz}} \and
        K.~Kampf\inst{1} \and
        J.~Novotn{\'y}\inst{1} 
}

\institute{Faculty of Mathematics and Physics, Institute of Particle and Nuclear Physics,
Charles University, V Holešovičkách 2, 18000 Praha 8, Czech Republic
          }

\abstract{
The radiative corrections to double-Dalitz ($P\to \bar{\ell}\ell \bar{\ell}'\ell' $) decays are revisited and completed up to next-to-leading order in QED, finding mild differences with respect to previous studies. These might be relevant for extracting information about the mesons transition form factors, which play an important role in determining the hadronic light-by-light contribution to the anomalous magnetic moment of the muon. 
}
\maketitle
\section{Introduction}
\label{intro}

Since the first measurements of the $\pi^0\to e^+e^-\gamma$ Dalitz decay in the early 60's~\cite{Tanabashi:2018oca}, there has been much interest in pseudoscalar mesons ($\pi^0, \eta, \eta'$) Dalitz decays which, despite the challenges, have seen tremendous experimental progress in the recent years~\cite{TheNA62:2016fhr,Adlarson:2016ykr,Arnaldi:2016pzu,Adlarson:2016hpp,Ablikim:2015wnx}. The relevance of these decays is related to the fact that they probe the mesons electromagnetic structure encoded in their transition form factors (TFFs), $F_{P\gamma^*\gamma^*}(q_1^2,q_2^2)$,

\begin{equation}
   \int d^4x e^{i q_1\cdot x} \langle 0| T\{ j_{\mu}(x) j_{\nu}(0) \} |P\rangle \equiv 
   -i\epsilon_{\mu\nu\rho\sigma}q_{1}^{\rho}q_{2}^{\sigma} F_{P\gamma^*\gamma^*}(q_1^2,q_2^2), 
\end{equation}
which are hard to predict theoretically from first principles---especially at the low energies probed in these processes. As such, they are valuable for testing and improving current models for the TFFs. 

At present, the relevance of such information can be understood from the key role that they play in determining the hadronic light-by-light (HLbL) contribution to the anomalous magnetic moment of the muon~\cite{Jegerlehner:2009ry,Masjuan:2017tvw}. However, Dalitz decays probe only the singly-virtual TFF, $F_{P\gamma^*\gamma^*}(q_1^2,0)$, while the doubly-virtual one, $F_{P\gamma^*\gamma^*}(q_1^2,q_2^2)$, is required for the HLbL as well. This gap could be closed by measuring the so called double-Dalitz decays ($P\to \bar{\ell}\ell \bar{\ell}'\ell' $), which are sensitive to the doubly-virtual TFF.

However, before extracting information about the TFFs from experiment, it is instrumental to keep control on QED radiative corrections, which can distort the hadronic effects (this is, the TFF). Very recently, these corrections were revisited for the Dalitz decays, finding important corrections with respect to previous studies~\cite{Husek:2015sma}. In this work~\cite{Kampf:2018wau}, we address the radiative corrections for double-Dalitz decays, revising and completing the work in Ref.~\cite{Barker:2002ib}.

\section{The radiative corrections}
\label{sec:rc}

\begin{figure}[h]
    \centering
    \includegraphics[width=0.7\textwidth]{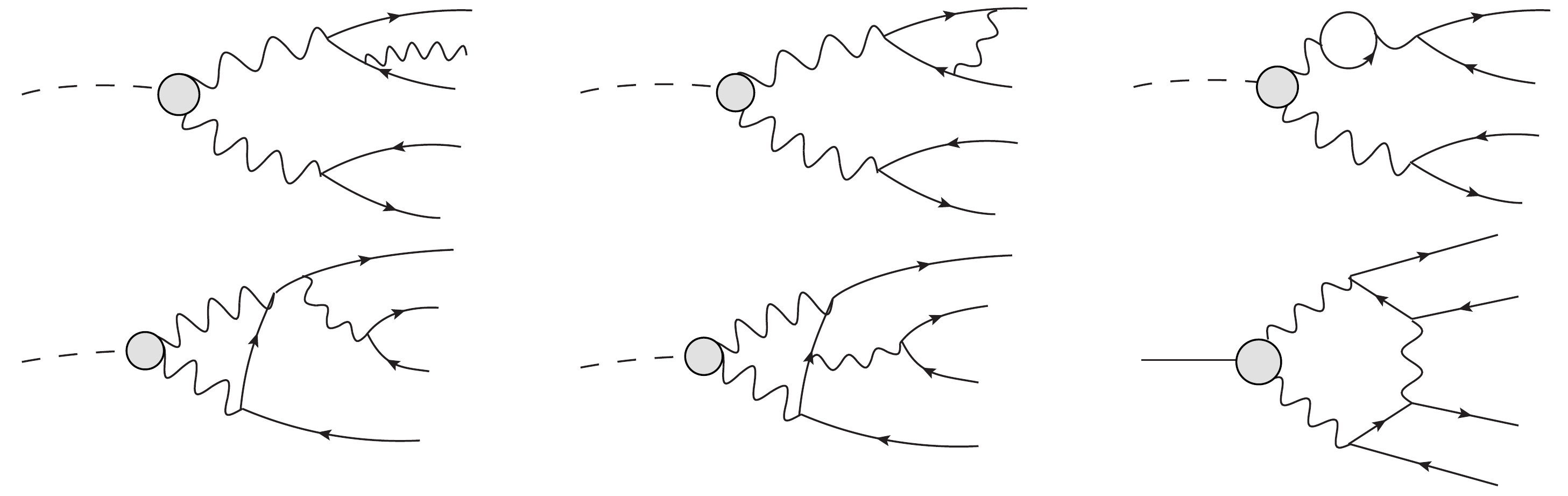}
    \caption{NLO representative diagrams. Upper row: bremsstrahlung, vertex, vacuum polarization. 
             Lower row: three-, four-, and five-point amplitudes.}
    \label{fig:nlo}
  \end{figure}

At leading order (LO), the double-Dalitz decay amplitude is given as 
\begin{equation}
    i\mathcal{M}^{\textrm{LO}} = -ie^4\frac{F_{P\gamma\gamma}(s_{12},s_{34})}{s_{12} s_{34}}
                                     \epsilon_{\mu\nu\rho\sigma}p_{12}^{\mu}p_{34}^{\rho} 
                                     \left[ \bar{u}(p_1)\gamma^{\nu}v(p_2)\right]  \left[\bar{u}(p_3)\gamma^{\sigma}v(p_4) \right],
\end{equation}
with an additional (exchange) contribution whenever identical leptons appear in the final state. Concerning the direct term only, this leads to 
\begin{equation}
    |\mathcal{M}^{\textrm{LO}}|^2 = \frac{e^8 |F_{P\gamma\gamma}(s_{12},s_{34})|^2}{x_{12}x_{34}}\lambda^2 
    \Big( 2 - \lambda_{12}^2 + y_{12}^2 -\lambda_{34}^2 + y_{34}^2 +(\lambda_{12}^2 - y_{12}^2)(\lambda_{34}^2 - y_{34}^2)\sin^2\phi\Big),
\end{equation}
where the role of the doubly-virtual TFF is clear (for definitions and the exchange amplitude, we refer to Ref.~\cite{Kampf:2018wau}). Concerning higher-order virtual corrections, we express the result as 
\begin{equation}\label{eq:MSQ}
 |\mathcal{M}|^2 = |\mathcal{M}^{\textrm{LO}} + \mathcal{M}^{\textrm{NLO}} + ...|^2 = 
                   |\mathcal{M}^{\textrm{LO}}|^2 + 2\operatorname{Re}\mathcal{M}^{\textrm{NLO}}\mathcal{M}^{\textrm{LO}*} + ...
                 \equiv \textrm{LO} + \textrm{NLO},
\end{equation}
with obvious meanings. In addition, infrared (IR) divergences appearing in loops are to be cancelled by real emission grahs (bremsstrahlung) that have to be added as well. Consequently, we divide the radiative corrections into bremsstrahlung, vertex (including self energies), vacuum polarization, three-, four-, and five-point ones, whose representative diagrams can be found in Fig.~\ref{fig:nlo} (see Ref.~\cite{Kampf:2018wau} for details).

Concerning bremsstrahlung (BS), we computed it in the soft-photon approximation:
\begin{equation}
  |\mathcal{M}^{\textrm{BS}}|^2 = e^2|\mathcal{M}^{\textrm{LO}}|^2 \sum\nolimits_{i,j} -\mathcal{Q}_i\mathcal{Q}_j I(p_i,p_j),
\end{equation}
where $p_i(\mathcal{Q}_i)$ refers to the $i$-th particle momentum(charge), and $I(p_i,p_j)$ is related to a well-known integral~\cite{tHooft:1978jhc,Kampf:2018wau}. We found agreement with Ref.~\cite{Barker:2002ib}, the exception being the $p_i=p_j$ cases which have, at least, an unclear definition. Concerning vertex corrections, and defining 
\begin{equation}
 \langle \ell(p') | j^{\mu} | \ell(p) \rangle = 
  \bar{u}(p)\left[ \gamma^{\mu}F_1(q^2) +i\frac{\sigma^{\mu\nu}}{2m_{\ell}}q_{\nu}F_2(q^2)  \right]v(p'); \quad q = p+p'.
\end{equation}
they shift the LO result, $F_{1(2)}(q^2)=1(0)$, to $q^2$-dependent form factors well-known at NLO~\cite{Kampf:2018wau} and in agreement with \cite{Barker:2002ib}. The $F_1$-correction is then trivial to compute, while the $F_2$ correction requires to evaluate a new matrix-element squared. From our analytic results, we find differences with respect to those in Ref.~\cite{Barker:2002ib} (find details in \cite{Kampf:2018wau}).\footnote{Also some ambiguities appear in \cite{Barker:2002ib} for direct and exchange interference terms, see Ref.~\cite{Kampf:2018wau}.}\\

Concerning the three- and four-point amplitudes, we computed them for the first time, with analytic results for direct terms given in terms of loop functions~\cite{Kampf:2018wau}. Numerically, this correction turns out to be similar in size to the $F_2$ vertex correction. 

Regarding the five-point amplitude, this was already computed in Ref.~\cite{Barker:2002ib}, as it is necessary to cancel IR divergences. Since analytic evaluations are too involved (with rank-3 five-point loop functions appearing), we provide analytic results for the amplitudes only as in \cite{Barker:2002ib} and with good agreement.
For its evaluation we employed two different methods. In the first, we first evaluated the spinor sums, that allowed to reduce everything down to scalar five-point functions and lower-point ones. In the second, we first expressed the amplitude in terms of tensor five-point functions, evaluating the spinor sums afterwards. The loop functions were evaluated with LoopTools~\cite{Hahn:1998yk} and agreement was found.\footnote{We crossed-checked the results of five-point loop functions by using the method in Ref.~\cite{Denner:2002ii}.} Remarkably, charge conjugation allows to show that the overall five-point amplitudes correction to the decay width vanishes except when identical particle appears, which can be used as a check of the numerics.

\section{Numerical results}
\label{sec:numerics}

Finally, we provide here the numerical result for the correction to the branching ratio (BR) in Table~\ref{tab:res},\footnote{For individual contributions' corrections, we refer to Ref.~\cite{Kampf:2018wau}.} which is given as $\delta(\textrm{NLO})= (\textrm{BR}^{\textrm{LO+NLO}}/\textrm{BR}^{\textrm{LO}}-1)$ (see notation in Eq.~\ref{eq:MSQ}).\footnote{We use, following Ref.~\cite{Barker:2002ib}, a soft-photon cutoff $x_{4\ell}=0.9985$ (see Ref.~\cite{Kampf:2018wau} for details).} For the numeric evaluation we used the CUBA library~\cite{Hahn:2004fe}. Moreover, we employed a logarithmic rescaling for invariant masses that improved the efficiency and might be useful in Monte Carlo (MC) generators. As said, we checked that the contribution of the five-point amplitude integrated to zero within errors for identical leptons. 
\begin{table}[!htb]
\centering
\caption{Our numeric result for the overall radiative corrections (first row). The second shows the effects of including a TFF on the LO result. The third row conatains the same corrections included in Ref.~\cite{Barker:2002ib} (in fourth row), that show the disagreement. The last row stands for the BR at NLO accuracy.}
\label{tab:res}
\resizebox{\textwidth}{!}{\small
  \begin{tabular}{cccccccc}\hline
                   & $\pi^0\to 4e$ & $K_L\to 4e$ & $K_L\to 2e2\mu$ & $K_L\to 4\mu$ & $\eta\to 4e$ & $\eta\to 2e2\mu$ & $\eta\to 4\mu$  \\ \hline
  $\delta(\textrm{NLO})$ &   $-0.1727(2)$ & $-0.2345(1)$ & $-0.0842(2)$ & $0.0608(2)$ & $-0.2409(1)$ & $-0.0900(1)$ & $0.0455(2)$ \\
  $\delta(\textrm{FF})$ &  $\phantom{-}0.0037(2)$ & $\phantom{-}0.0749(2)$ & $\phantom{-}0.6942(2)$ & $0.8608(3)$ & $\phantom{-}0.0207(2)$ & $\phantom{-}0.4829(2)$ & $0.6202(3)$ \\ \hline
  no 3,4       &$-0.1718(2)$ &$-0.2262(2)$ & $-0.0767(1)$ & $0.0704(1)$ & $-0.2301(1)$ & $-0.0836(1)$ & $0.0535(1)$ \\ 
  Barker              &$-0.160(2)$  &$-0.218(1)$  & $-0.066(1)$ & $0.084(1)$ & $-$ & $-$ & $-$ \\  \hline
  BR(LO+NLO)              &$2.840(1)10^{-5}$  &$5.120(1)10^{-5}$  & $4.436(1)10^{-6}$ & $1.851(1)10^{-9}$ & $5.202(1)10^{-5}$ & $5.393(1)10^{-6}$ & $10.289(2)10^{-9}$ \\  \hline
  \end{tabular}
}
\end{table}

Comparing to Ref.~\cite{Barker:2002ib}, we find slight different numerics even when three- and four-point amplitudes are ommitted. However, the differences cannot be attributed to differences in the $F_2$ correction or bremsstrahlung---the former is too small and the latter increases the difference. One possibility could be problems in ther five-point amplitude or numerics in its evaluation. In any case, the differences we find are relevant if one is willing to extract any information about the doubly-virtual TFFs.

\section{Summary and Outlook}

In our work, we have re-evaluated and completed the NLO QED radiative corrections for double-Dalitz decays in the soft-photon approximation. Comparing to the previous work~\cite{Barker:2002ib}, we have found some differences with respect to their values, which origin we could not however trace back. Concerning the contributions ommitted in the previous study, we find them small, but of the size of $F_2$ vertex corrections. Overall, the difference we find seems relevant for extracting information about the TFFs. This might be possible for instance at REDTOP experiment~\cite{Gatto:2016rae}, which expects to produce of the order of $10^{12(11)}$ $\eta(\eta')$ mesons.

\begin{figure}[t]
  \centering
    \includegraphics[width=\textwidth]{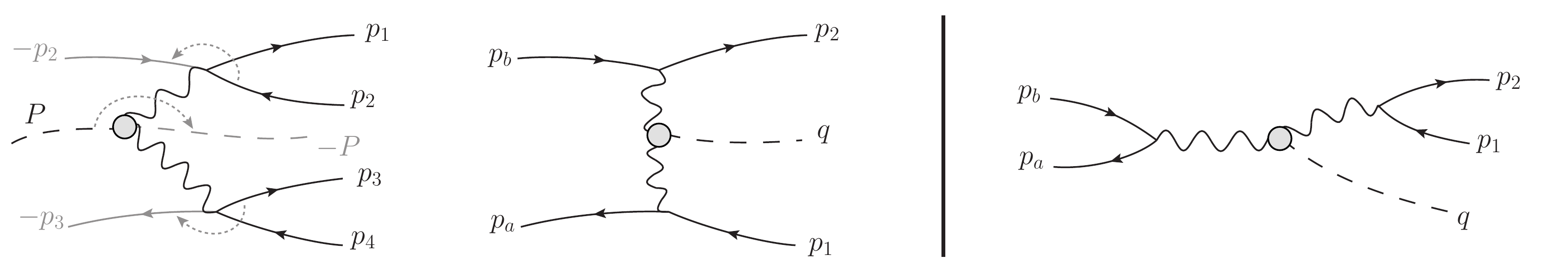}
  \caption{Left block shows the choice for crossing relations from direct double-Dalitz terms (left) to the $e^+e^- \to e^+e^-P$ 
           $t$-channel ones (center). The right block is the s-channel contribution to the latter process that would be crossing-related 
           to the double-Dalitz exchange diagrams.\label{fig:crossing}}
  \end{figure}

As a future work, our results can be connected through crossing symmetry to the $e^+e^- \to e^+e^-P$ process, see Fig.~\ref{fig:crossing}.\footnote{At the amplitude level this is realized through $p_1 \to p_2$, $p_2 \to -p_b$, $p_3 \to -p_a$, $p_4 \to p_1$, $P \to -q$,    $\bar{u}_1 \to \bar{u}_2$, $v_2 \to u_b$, $\bar{u}_3 \to \bar{v}_a$, $v_4\to v_1$, where we follow the notation in Ref.~\cite{Schuler:1997ex} (see also Ref.~\cite{Czyz:2010sp}).}
Such corrections are relevant for next-generation MC, such as Ekhara~\cite{Czyz:2010sp}---a common tool for experimentalists. In its latest
version~\cite{Czyz:2018jpp} all but the $n$-point amplitudes (and corresponding IR-divergent bremsstrahlung processes) have been included. Therefore, there is a current joint effort in implementing the five-point functions into Ekhara, which seems the most relevant given their role in cancelling IR-divergences~\cite{inprep}.

\begin{acknowledgement}
This work was supported by the Czech Science Foundation (Grant No. GACR 18-17224S) and by the project UNCE/SCI/013 of Charles University.
\end{acknowledgement}

%
\bibliography{references}
%
%

\end{document}